\documentclass[aps,floats,floatfix,amssymb,prd,
onecolumn,superscriptaddress,nofootinbib,notitlepage,11pt]{revtex4-1}

\usepackage{slashed}
\usepackage{bm}
\usepackage{latexsym,amssymb,amsmath,float,url,mathrsfs}
\usepackage{latexsym}
\usepackage{graphicx}
\usepackage{epstopdf}
\usepackage{amsmath}
\usepackage{comment}
\usepackage{subfigure}
\usepackage{natbib}
\usepackage{hyperref}
\usepackage{pifont}
\usepackage{blindtext}
\usepackage{adjustbox}
\usepackage{multirow}
\usepackage{tabularx}
\usepackage[table]{xcolor}
\usepackage{orcidlink}
\usepackage{tikz}
\usetikzlibrary{tikzmark}

\usepackage{blkarray}
\usepackage{graphicx}
\usepackage{amsfonts}
\usepackage{soul}
\usepackage{amssymb}
\usepackage{amsmath}
\usepackage{cancel}
\usepackage{tcolorbox}

\usepackage{graphics,appendix,afterpage,makecell} 

\definecolor{oucrimsonred}{rgb}{0.6, 0.0, 0.0}
\definecolor{persianblue}{rgb}{0.11, 0.22, 0.73}
\definecolor{forestgreen}{rgb}{0.13,0.35,0.13}
\definecolor{lightgray}{rgb}{0.83, 0.83, 0.83}
 \hypersetup{colorlinks, citecolor=oucrimsonred, linkcolor=black, urlcolor=oucrimsonred}
\definecolor{cornellred}{rgb}{0.7, 0.11, 0.11}
\definecolor{navyblue}{rgb}{0.0, 0.0, 0.5}
\definecolor{amethyst}{rgb}{0.6, 0.4, 0.8}
\definecolor{yellow}{rgb}{1.0, 1.0, 0.0}
\definecolor{firebrick}{rgb}{0.7, 0.13, 0.13}
\definecolor{tangerineyellow}{rgb}{1.0, 0.8, 0.0}
\definecolor{deepfuchsia}{rgb}{0.76, 0.33, 0.76}
\definecolor{amber}{rgb}{1.0, 0.75, 0.0}
\definecolor{VioletRed4}{rgb}{0.55, 0.13, .32}
\definecolor{indiagreen}{rgb}{0.07, 0.53, 0.03}
\definecolor{VioletRed4}{rgb}{0.55, 0.13, .32}
\newcommand{\be}{\begin{equation}}
\newcommand{\ee}{\end{equation}}
\newcommand{\bea}{\begin{equation} \begin{aligned}}
\newcommand{\eea}{\end{aligned} \end{equation}}

\definecolor{oucrimsonred}{rgb}{0.6, 0.0, 0.0}
\newcommand\vertarrowbox[3][6ex]{%
  \begin{array}[t]{@{}c@{}} #2 \\
  \left\uparrow\vcenter{\hrule height #1}\right.\kern-\nulldelimiterspace\\
  \makebox[0pt]{\scriptsize#3}
  \end{array}%
}

\definecolor{verdechiaro}{rgb}{0.6,1,0.6}
\definecolor{giallochiaro}{rgb}{1,1,0.6}
\definecolor{bluscuro}{rgb}{0.15, 0.2, 0.9}
\definecolor{verdes}{rgb}{0.1, 0.5, 0.1}%
\definecolor{tangerineyellow}{rgb}{1.0, 0.8, 0.0}

\definecolor{americanrose}{rgb}{1.0, 0.01, 0.24}
\definecolor{cobalt}{rgb}{0.0, 0.28, 0.67}
\definecolor{brandeisblue}{rgb}{0.0, 0.44, 1.0}
\definecolor{mycolor}{rgb}{0.0, 0.0, 0.5}
\definecolor{oxfordblue}{rgb}{0.0, 0.13, 0.28}
\definecolor{azure}{rgb}{0.0, 0.5, 1.0}
\definecolor{turquoiseblue}{rgb}{0.0, 1.0, 0.94}
\newtcolorbox{mynewbox}[1]{colback=white!5!white,colframe=azure!75!black,fonttitle=\bfseries,title=#1}
\newtcolorbox{mybox}{colback=mycolor!5!white,colframe=azure!75!black}
\newtcolorbox{mynamedbox}[1]{colback=mycolor!5!white,colframe=azure!75!black,title=#1}
\definecolor{venetianred}{rgb}{0.78, 0.03, 0.08}
\newtcolorbox{mynamedbox1}[1]{colback=venetianred!5!white,colframe=venetianred!80!black,title=#1}
\newtcolorbox{mynamedbox2}[1]{colback=azure!5!white,colframe=azure!80!black,title=#1}

\definecolor{verdes}{rgb}{0.1, 0.5, 0.1}%
\definecolor{cornellred}{rgb}{0.7, 0.11, 0.11}

\definecolor{VioletRed4}{rgb}{0.55, 0.13, .32}

\hypersetup{
     colorlinks   = true,
     citecolor    = violet,
     urlcolor     = violet,
     linkcolor    = violet}

\definecolor{rossocorsa}{rgb}{0.83, 0.0, 0.0}

\usepackage[normalem]{ulem}

\newcommand{\papertitle}{Why the Universal   Threshold for  Primordial Black Hole   Formation is Universal}

\begin{document}

\title[]{\papertitle}

\author{Alex Kehagias\orcidlink{}}
\affiliation{Physics Division, National Technical University of Athens, Athens, 15780, Greece}

\author{Davide Perrone\orcidlink{0000-0003-4430-4914}}
\affiliation{Department of Theoretical Physics and Gravitational Wave Science Center,  \\
24 quai E. Ansermet, CH-1211 Geneva 4, Switzerland}

\author{Antonio Riotto\orcidlink{0000-0001-6948-0856}}
\affiliation{Department of Theoretical Physics and Gravitational Wave Science Center,  \\
24 quai E. Ansermet, CH-1211 Geneva 4, Switzerland}


\begin{abstract}
\noindent
We show why the threshold for primordial black hole formation is   universal  (independent from  the shape of the  perturbation) when expressed in terms of the volume averaged compaction function. The proof is rooted  in  the  
self-similarity  of the  gravitational collapse phenomenon at criticality.

\end{abstract}

\maketitle

\section*{1. Introduction}
\noindent
The topic of Primordial Black Holes (PBHs) has become much debated  in the last years (see Ref. \cite{LISACosmologyWorkingGroup:2023njw} for a recent review) since they  might provide an explanation for  some of the signals from binary BH  mergers measured by gravitational wave detectors and  an important component of the dark matter in the universe. Furthermore, the  
 next generation of gravitational wave experiments will provide an armoury of  smoking guns to  distinguish PBHs from their astrophysical counterparts at high redshifts \cite{Riotto:2024ayo}.

The most common scenario for the formation of the PBHs is the one where they are originated  from  the collapse of large fluctuations formed during inflation upon horizon re-entry \cite{Sasaki:2018dmp}. The formation probability, and therefore their current abundance,  is extremely  sensitive to the critical threshold of collapse, that is on the minimum value ${\cal C}_c$ that   the amplitude of the compaction function (or the energy overdensity) must have in order for a fluctuation to form a BH \cite{Yoo:2022mzl}.

Numerical simulations  have shown that  
the critical threshold depends sensitively on the curvature at the peak of the compaction function \cite{Musco:2018rwt, Germani:2018jgr,Escriva:2019phb,Musco:2020jjb}, that is on the shape of the perturbation.   The threshold of the compaction function varies from 2/5 for broad profiles to 2/3 for very peaked profiles. This     is certainly not a pleasant feature   since 
 the theory delivers only stochastic quantities and one can only     calculate   the  averaged profile of a perturbation. Furthermore, in  realistic cases  the critical threshold for formation is determined by the broadest possible compaction function, and not by the mean profile.
\cite{Ianniccari:2024bkh}. 

This ambiguity can be eliminated by the    interesting observation  made in Ref. \cite{Escriva:2019phb}. Based on a numerical approach,  it has been shown that the threshold to form PBHs from an initial spherically symmetric perturbation is universal when   the compaction function is averaged over a sphere of radius equal to the location of the maximum of the compaction function. Its critical value in radiation turns out to be  2/5 independently from the shape of the perturbation and within the numerical errors \cite{Escriva:2019phb}.

The goal of this short paper is to provide a simple proof  of why such universal threshold for the formation of PBH is indeed universal and independent of the shape of the collapsing compaction function. As we shall see, our proof is  rooted in the fundamental property of the phenomenon of gravitational collapse: 
the existence of  attractors within the BH threshold, that is, attractors of codimension one in phase space, and which are self-similar. More specifically, 
the persistence of   self-similarity  at criticality will play a key role.

\vskip 0.5cm
\section*{2. Some basic concepts}
\noindent
PBH formation is routinely studied for  spherical and time-dependent configurations.  Therefore, we  will be concerned  with time-dependent spherically symmetric spacetimes. The latter have a residual symmetry which is more manifest if the  metric   is  written locally in double-null form as (see, for instance, Ref. \cite{Hayward:1994bu})

\be
\label{null}
{\rm d}s^2=-2e^{-f}{\rm d}\xi^+{\rm d}\xi^- +R^2{\rm d}\Omega^2,
\ee
where ${\rm d}\Omega^2$ refers to the unit sphere and $R$ and $f$ are functions of the null coordinates $\xi^\pm$.   In such double-null form  the residual coordinate freedom consists  of all the  diffeomorphisms

\be
\label{res}
\xi^\pm\rightarrow\widetilde{\xi}^\pm(\xi^\pm).
\ee
The key point is that the areal radius $R$ is a geometrical invariant quantity (even if $f$ is not) under the residual symmetry, that is it transforms like a scalar (see, for instance, Ref. \cite{Faraoni:2016xgy}).
One could also use one spatial and one temporal direction (which we will adopt in the following) to show it,  but there is no unique choice of such directions and this  makes it more difficult to check coordinate invariance. 

The Misner-Sharp spherically symmetric gravitational energy (in units $G_N = 1$ and $a=\pm$) \cite{Misner:1964je} 

\begin{eqnarray}
M&=&\frac{1}{2}R\left(1-\nabla_a R\,\nabla^a R\right)\nonumber\\
&=&\frac{1}{2}R+e^f \,R\,\partial_+R\,\partial_- R
\end{eqnarray}
is consequently an invariant.
Finally, we define the  compaction function, which  we will use   to  set  the criterion for the PBH formation, to be  

\be
\label{comp}
C=\frac{2 M}{R}.
\ee
Being constructed from the Misner-Sharp mass and the areal radius, it is as well a  geometric quantity   
independent from any spherically symmetric foliation.

The  rescaling freedom (\ref{res}) can be used to fix the null coordinates

\be
\sqrt{2} e^{-f/2}{\rm d}\xi^\pm=A{\rm d}t\pm B{\rm d}r,
\ee
such that the 
 metric (\ref{null}) becomes

\be
\label{metric}
{\rm d}s^2=-A^2{\rm d}t^2+B^2{\rm d}r^2+R^2{\rm d}\Omega^2,
\ee
where now $R$, $A$  and $B$ are functions of $(r,t)$. Denoting the derivatives with respect to $r$ with a prime and the derivative with respect to $t$ with a dot, one can write the equations of motion for the Misner-Sharp mass  given a perfect fluid  with energy density $\rho$ and pressure density $P$ \cite{Misner:1964je}

\begin{eqnarray}
M'&=&4\pi R^2\rho R',\nonumber\\
\dot M&=&-4\pi R^2 P \dot R.
\end{eqnarray}
Near  a regular center $R=0$, one has $R'={\cal O}(R^0)$ and, if $\partial/\partial t$ is tangent to the centre, $\dot R={\cal O}(R)$. As a consequence, if $\rho$ is ${\cal O}(R^0)$, then one can conclude in full generality 
that \cite{Harada:2001nh}

\be
\label{lim}
M(r\sim 0,t)=\frac{4\pi}{3} \rho(r\sim 0,t)R^3(r\sim 0,t)+{\cal O}(R^4),
\ee
for any given time. It is important to notice that this behavior  is completely general and gauge-independent. It does not require the solution to be self-similar, but it only depends on the regularity of the solution at the origin.
We will use this fundamental result in the following.

\vskip 0.5cm
\section*{3. The self-similarity at criticality}
\noindent
The first description of critical phenomena in gravitational collapse goes back to the seminal references \cite{Choptuik:1993,Evans:1994}. The 
BH threshold in the space of initial data for general relativity shows a surprising simple structure characterized by  universality and   power-law scaling of the BH mass (for a review, see Ref. \cite{Gundlach:2002sx}).  This is  explained by the existence of exact solutions which are self-similar attractors near the center of collapse.

The dynamics is the following. A given overdensity of the radiation fluid has an initial comoving scale larger than the comoving Hubble radius. At some point in time the  latter catches the size of the overdensity and it is at this moment that numerical simulations provide the threshold value for PBH formation. It is also the moment when the metric and the fluid quickly approach a self-similar behaviour \cite{Musco:2012au} which depends only on the variable 
\be
z=\frac{r}{(-t)},\,\,\,\,\,\,t<0
\ee
and independent from the time variable 
\be
\tau=-\ln(-t).
\ee
At later times, self-similarity is broken, leading eventually to the formation of a BH if the evolution is super-critical, that is if the compaction function at its maximum is larger than a critical value (for a review, see Ref. \cite{LISACosmologyWorkingGroup:2023njw}).

The key point for us is that near,  but not precisely critical,  spacetimes develop a self-similar region for $r$ much smaller than the
initial size of the configuration, but then exit  self-similarity on even smaller, but finite  scales, in order to form a BH. 

On the contrary,  
the critical configuration for  BH formation is the one that approaches  the self-similar solution upon horizon re-entry and retains the self-similar  dependence even when  $r\rightarrow 0$ and $t\rightarrow 0$. In other words,  by definition,  critical configurations out of which one deduces the threshold for BH formation, are defined to be those that stay {\it always} self-similar. 

It is this property of never-ending self-similarity for the critical configurations which we will use in the next section to define the most suitable threshold for PBH formation. 

\vskip 0.5cm
\section*{4. Why the universal threshold for PBH formation is universal}
\noindent
Let us consider a PBH formation during the radiation phase of the early universe.
We define  the volume averaged compaction function at a given  time $t_0$ as
\begin{eqnarray}
\overline{C}(R)=\frac{3}{R^3}\int_0^R {\rm d}x \,x^2\,C(x), \label{AC}
\end{eqnarray}
where $C(R(r,t_0))=C(r,t_0)$. The time $t_0$ has to  be thought of as the time a given perturbation enters the horizon and the self-similar behaviour takes over.
The quantity (\ref{AC}) is manifestly gauge-invariant being constructed from gauge-invariant quantities. 

We now   rescale the spatial and time coordinates performing a dilatation transformation 

\be
\label{dil}
r\to \lambda r\,\,\,\,{\rm and}\,\,\,\,  t\to \lambda t,
\ee
such that the variable $z$ is invariant
\be
z\to z.
\ee
Since the spacetime is self-similar,  it admits a conformal Killing vector, which for the metric (\ref{metric}) is given by
\begin{eqnarray}
    \xi=t\partial_t+r \partial_r, 
\end{eqnarray}
and satisfies 
\begin{equation}
    \nabla_\mu\xi_\nu+\nabla_\nu \xi_\mu=2g_{\mu\nu}. 
\end{equation}
It is easy to see that the compaction function defined in Eq. (\ref{comp}) is invariant under the dilatation (\ref{dil}) 
\be
\delta C=t\partial_t C+r\partial_r C=0, \label{ccc1}
\ee
and depends only on the zooming coordinate $z$. Similarly, the averaged compaction function is also invariant since 
\be
\delta \overline{C}(R)=3C(R)-3\overline{C}(R)+\frac{3}{R^3}
\int_0^R {\rm d}x \,x^2\,t\partial_t C(x),
\ee
and thus, after using Eq. (\ref{ccc1}), one finds 
\be
\delta \overline{C}(R)=0.
\ee
This implies that one can calculate $\overline{C}(R)$ at any $r$, and in particular around $r=0$, by appropriate rescalings of both $r$ and $t$.
Since the areal radius $R$ scales like $r$ (it can be always be expressed as $r$ times a function of $z$ \cite{Harada:2001nh}), that is 
\be
R\to \lambda R,
\ee
then the volume averaged compaction function 
 scales as 
\begin{eqnarray}
\overline{C}(R)&=&\frac{3}{(\lambda R)^3}\int_0^{\lambda R}{\rm d}(\lambda x) \,({\lambda x})^2\,C(\lambda x) \nonumber\\
&=&\frac{3}{\overline R^3}\int_0^{\overline R}{\rm d}\overline x \,\overline x^2\,C(\overline x),\label{AC1}
\end{eqnarray}
where $\overline{R}=\lambda R$. 

The key point is that we can now  take $\lambda$ as small as we wish in order to   bring $\lambda R$ as close as desired to the central origin. For what explained in the previous section,  for the critical configuration the self-similar solution persists in this limit independently from how close we are at $R\sim r\sim 0$.

We can use Eqs.  (\ref{lim}) and (\ref{comp}) to express Eq. (\ref{AC1}) as 

\begin{eqnarray}
\overline{C}(R)&=&\frac{3}{\overline R^3}\int_0^{\overline R}{\rm d}\overline x\,
\frac{8\pi}{3}\,\overline{x}^4\,\rho(0,\lambda t_0)\,
 \nonumber \\
&=&\frac{3}{\overline R^3}\int_0^{\overline R}
{\rm d}\overline x \frac{8\pi}{3}\lambda^{-2}\overline{x}^4 \rho(0,t_0), \label{AC2}
\end{eqnarray}
where we have used  the scaling  
$
\rho(0,\lambda t_0)= \lambda^{-2}\rho(0, t_0)$, which comes from the fact that in full generality $t^2\rho(0,t)=$ constant \cite{Harada:2001nh}.
From Eq. (\ref{AC2}) we find  that  
\begin{eqnarray}
\overline{C}(R)&=&\frac{3}{5} \left(\frac{8\pi}{3}\rho(0,t_0)
 R^2(0,t_0)\right)\nonumber \\
 &=& \frac{3}{5}\left(\lim_{r\to 0}\frac{2M(r,t_0)}{R(r,t_0)}\right)\nonumber\\
 &=&\frac{3}{5},
\end{eqnarray}
where in the last passage we have used the fact that for critical solution the formation of the BH imposes 
\begin{eqnarray}
\lim_{r\to 0}\frac{2M(t_0,r)}{R(t_0,r)}= 1
\end{eqnarray}
since  the critical BH has zero mass and therefore zero horizon radius.
In general for a fluid of equation of state $\omega$ one has

\begin{eqnarray}
\overline{\mathcal{C}}=f(\omega)\overline{C},\,\,\,f(\omega)=\frac{3(1+\omega)}{5+3\omega},
\end{eqnarray}
such that for radiation ($\omega=1/3$) we finally obtain

\begin{eqnarray}
\overline{\mathcal{C}}_c=\frac{2}{5}.
\end{eqnarray}
 This is the value obtained by numerical simulations. 
 We have thus proven that the volume averaged compaction function provides a universal threshold for the PBH formation because  the critical solution exhibits persistent self-similarity  near the center
 which is rooted in the critical behavior of the gravitational collapse phenomenon. It will be interesting to understand if our result has an interepretation in view of the correspondence between the formation of PBHs and the stability of circular null geodesics around the collapsing perturbation 
\cite{Ianniccari:2024ltb}, where self-similarity plays also a crucial role.


\vskip 0.5cm

\begin{acknowledgments}
\noindent
 We thank V. De Luca for useful comments. A.R.  acknowledges support from the  Swiss National Science Foundation (project number CRSII5\_213497) and from 
the Boninchi Foundation for the project ``PBHs in the Era of GW Astronomy''.
\end{acknowledgments}

\bibliography{main}
\end{document}